# Anisotropic Ginzburg-Landau scaling of $H_{c2}$ and transport properties of 112-type $Ca_{0.8}La_{0.2}Fe_{0.98}Co_{0.02}As_2$ single crystal

Xiangzhuo Xing[1], Wei Zhou[1], Nan Zhou[1], Feifei Yuan[1], Yongqiang Pan[1], Haijun Zhao[1], Xiaofeng Xu[2] and Zhixiang Shi[1]

[1]*Department of Physics and Key Laboratory of MEMS of the Ministry of Education, Southeast University, Nanjing 211189, China*

[2]*Department of Physics and Hangzhou Key Laboratory of Quantum Matters, Hangzhou Normal University, Hangzhou 310036, China*

E-mail: zxshi@seu.edu.cn

## Abstract

High-quality single crystal of $Ca_{0.8}La_{0.2}Fe_{0.98}Co_{0.02}As_2$ has been successfully synthesized using a self-flux method. The magnetization measurement reveals a second peak effect and high critical current density $J_c$ exceeding $2 \times 10^6$ A/cm$^2$ at 5 K (self-field). The upper critical field anisotropy was systematically studied by measuring the electrical resistivity under various magnetic fields and angles. The angle-dependent magnetoresistance, by choosing an appropriate anisotropy parameter within the framework of the anisotropic Ginzburg-Landau (AGL) theory, can be scaled onto one single curve. In the normal state, the negative Hall coefficient shows strong but nonmonotonic $T$-dependence through a minimum at ~175 K. Moreover, it is shown that the magnetoresistance apparently violates the semiclassical Kohler's rule below ~ 175 K but can be well scaled by the Hall angle instead. This suggests either the change of carriers with $T$ or the exotic anisotropic scattering in the system.

## 1. Introduction

Since the discovery of superconductivity at 26 K in $LaFeAsO_{1-x}F_x$[1], a wide range of iron based superconductors (IBSs) have been discovered, such as $AeFe_2As_2$(Ae=alkaline earth,122-type) [2], LiFeAs(111-type)[3] and Fe(Se, Te)(11-type)[4]. Much attention has been paid to this new generation of high-temperature superconductors soon after. The fundamental parameters, such as the upper critical field $H_{c2}(T)$ and its anisotropy have been broadly investigated as they can provide valuable information on the coherent length, effective electron mass, superconducting pair-breaking mechanism, and in auspicious circumstances, can be harnessed for applications. For example, some IBSs show high upper critical field with $H_{c2}(0)$~100-200 T[5] and rather low upper critical field anisotropy compared with cuprate superconductors[6], suggesting promising prospect for applications. Usually, IBSs show pronounced $T$-dependence in the anisotropy of $H_{ci}$ due to the multiband features similar with $MgB_2$[7-10]. On the other hand, understanding the normal state properties is the key step to studying the electronic properties of a superconductor, which is intimately related to the superconducting mechanism. Indeed, numerous unusual behaviors in the normal state have been observed in IBSs, such as the non-Fermi liquid like $T$-linear resistivity[11,12], strong $T$-dependent Hall coefficients[13-15]. However, the interpretations remain controversial as to the importance of magnetic fluctuations or multiband effects[11,12]. Therefore, revealing the normal-state properties of IBSs is of crucial importance to understand their superconducting mechanisms.

Recently, a novel type of IBSs has been reported in $(Ca,RE)FeAs_2$(RE=Rare earth,112-type) with $T_c$ as high as 40 K[16-18]. $(Ca,RE)FeAs_2$ is crystallized in a monoclinic crystal structure with space group of $P21$ or $P21/m$, consisting of alternating stacking of FeAs and As zigzag bond layers. It immediately attracted tremendous interest due to its peculiar properties compared with other IBSs. For example, Density-functional theory calculations and angle-resolved photoemission spectroscopy measurements both identified that one additional hole-like band existed in the vicinity of $E_F$ and was absent in other iron-based superconductors[19,20]. In addition, the unique arsenic zigzag bond layers could generate anisotropic Dirac cones near the Fermi level[20]. Theoretical calculations further predict that $CaFeAs_2$ is a staggered intercalation compound that integrates both quantum spin Hall and superconductivity and may be an ideal system for the realization of Majorana related physics[21]. The discovery of $(Ca,RE)FeAs_2$ therefore provides a new platform to study the superconducting mechanisms of IBSs. Unfortunately, due to the limitation of the quality of single crystals, such as the broad superconducting transition and inhomogeneity, almost no systematical investigations have been made thus far to this new IBSs. Recently, Co doping in $Ca_{1-x}La_xFeAs_2$ has been used to improve superconducting properties in 112 system[22]. We have significantly improved the quality of single crystals by Co-co-doping method[23], and it becomes feasible to probe the intrinsic superconducting and normal state properties in a more detailed way. In this paper, we synthesized and characterized the high-quality $Ca_{0.8}La_{0.2}Fe_{0.98}Co_{0.02}As_2$ single crystals and systematically studied the upper critical field and its anisotropy. The normal state transport was also studied by the Hall effect and magnetoresistance (MR) measurements. Our data provides more detailed information on the electronic behavior of this new type IBSs.

## 2. Experimental details

High-quality single crystals with a nominal composition $Ca_{0.9}La_{0.1}Fe_{0.975}Co_{0.025}As_2$ were grown by the self-flux method as reported elsewhere[23,24]. Stoichiometric mixtures of Ca, La, FeAs, CoAs and As were placed in an aluminum crucible and subsequently sealed in an evacuated quartz tube. All the weighing and mixing procedures were carried out in an argon-filled glove box with the water and oxygen content below 0.1 ppm. The mixture was slowly heated up to 1100 ℃, held for 15 h to ensure the homogenization, and slowly cooled down to 1050 ℃ at a rate of 1.25 ℃/h. Then the temperature was cooled down to room temperature by switching off the furnace. Large single crystals with a typical dimension of 2×1×0.05 mm$^3$ were obtained. The single crystal X-ray diffraction measurement was performed on a commercial Rigaku diffractometer with Cu Kα radiation. Elemental analysis was performed by a scanning electron microscope equipped with energy dispersive X-ray (EDX) spectroscopy probe. Magnetic measurements were measured via the VSM option of a Quantum Design Physical Property Measurement System (PPMS). The electrical transport measurements were also performed on PPMS.

## 3. Results and discussion

The top inset of Fig.1 shows the single crystal XRD pattern of $Ca_{0.8}La_{0.2}Fe_{0.98}Co_{0.02}As_2$. Only (00$l$) diffraction peaks were observed, indicating good $c$-axis orientation. The actual stoichiometry of the single crystal determined by EDX analyses is $Ca_{0.8}La_{0.2}Fe_{0.98}Co_{0.02}As_2$. Similar with other reports, the actual content of La is higher than the nominal composition[22-25].The temperature dependence of resistivity was shown in the main panel of Fig. 1.The resistivity drops abruptly as the temperature decrease to $T_c^{onset}(90\%\rho_n)$=39.7 K and reaches zero at $T_c^{zero} = 37.8$ K with a transition width of ~1.6 K ($90\%\rho_n - 10\%\rho_n$), which indicates high quality of the sample. The bulk superconductivity of $Ca_{0.8}La_{0.2}Fe_{0.98}Co_{0.02}As_2$ is also confirmed by magnetization measurements shown in the bottom inset of Fig. 1. The sample shows sharp superconducting transition at $T_c$=37.8 K, consistent with the $T_c^{zero}$ obtained from the resistivity data.

Fig. 2(a) shows the magnetization hysteresis loops (MHLs) at several temperatures from 5 to 36 K in magnetic fields $H \parallel c$. The MHLs are big and symmetric, suggestive of the bulk vortex pinning. Moreover, the second-peak (SP) effect, was observed for the first time in this 112 system. The SP moves to higher fields with decreasing temperature and finally outreaches the available field range. Systematical studies on the SP effect in this system have been published in our recent paper [26]. We calculated the field dependence of critical current density $J_c$ from the MHLs based on the Bean Model[27] $J_c = 20\Delta M /[a(1-a/3b)]$, where $\Delta M$ is $M_{down} - M_{up}$, $M_{up}$ [emu/cm$^3$] and $M_{down}$ [emu/cm$^3$] are the magnetization when sweeping fields up and down, respectively, $a$ [cm] and $b$ [cm] are sample widths ($a < b$). Fig. 2(b) shows the field dependence of $J_c$. The calculated $J_c$ at 5 K and 0 T reaches $2\times10^6$ A/cm$^2$, larger than the Co-free samples and previous reported in Co doped (Ca,La)FeAs$_2$ polycrystalline samples[22,24]. Meanwhile, in comparison with other iron based superconductors, $J_c$ values extracted from magnetization measurement were summarized in Table I. The large $J_c$ proves the bulk superconductivity and suggests the potential applications in this system.

The $T$-dependent resistivity of the $Ca_{0.8}La_{0.2}Fe_{0.98}Co_{0.02}As_2$ single crystal in magnetic fields up to 9 T for $H\parallel ab$ and $H\parallel c$ configurations is shown in Fig. 3(a) and (b). The superconducting transition gradually shifts to lower temperatures and the width becomes significantly broadened with increasing field, which indicates the presence of strong thermal fluctuations of vortices. Fig. 3(c) shows the temperature dependence of $H_{c2}$ for both $H\parallel c$ and $H\parallel ab$, respectively. The values of $H_{c2}(T)$ were defined as the field at which 50% of the normal state resistivity reaches. $H_{c2}(T)$ exhibits a pronounced upward curvature, which is very different from the one-gap WHH behavior[32]. It may be attributed to the strong thermal fluctuations existed in high temperature superconductors.

Note the uncertainty for determining the anisotropy $\Gamma$ from the ratio of $H_{c2}^{ab}$ and $H_{c2}^{c}$, which is sensitive to the different criteria used in $H_{c2}$ determination[33-35]. We measured the angular dependence of resistivity under different magnetic fields at fixed temperatures in order to obtain the $\Gamma$ values more accurately. The insets of Fig. 4 present four sets of data at temperatures of 34 K, 35 K, 36 K and 37 K. All the curves show a cup-like shape with the minimum at $\theta = 90°$ ($H\|ab$) and maximum at $\theta = 0°$ and $180°$ ($H\|c$), where $\theta$ is the angle between the direction of the applied field and the $c$ axis. According to the anisotropic Ginzburg-Landau (AGL) theory[36], the effective upper critical field $H_{c2}^{GL}(\theta)$ at an angle $\theta$ can be obtained as $H_{c2}^{GL}(\theta) = H_{c2}^{ab}/\sqrt{\sin^2(\theta) + \Gamma^2\cos^2(\theta)}$, Where $\Gamma$ is the anisotropy parameter of the sample, $\Gamma = H_{c2}^{ab}/H_{c2}^{c} = (m_c/m_{ab})^{1/2} = \xi_{ab}/\xi_c$. Since the resistivity in the mixed state depends on the effective field $H/H_{c2}^{GL}(\theta)$, the resistivity can be scaled with $H/H_{c2}^{GL}(\theta)$. Thus using the scaling variable $\tilde{H} = H\sqrt{\sin^2(\theta) + \Gamma^2\cos^2(\theta)}$, the resistivity measured in different magnetic fields at a certain temperature should collapse onto one curve when an appropriate $\Gamma$ value is chosen. The results of this scaling at T=34 K, 35 K, 36 K, and 37 K are shown in the main panels of Fig.4. Excellent scaling can be obtained by adjusting $\Gamma$. The upper critical field anisotropy parameter estimated from $\Gamma = H_{c2}^{ab}/H_{c2}^{c}$ was also presented. It is found that the anisotropy parameter $\Gamma$ shows nonmonotonic dependence on $T$, first increases and then decreases slightly with decreasing temperature as plotted in Fig.5. In a large temperature region from 0.87 $T_c$ to 0.97 $T_c$, the $\Gamma$ values are located in 4~5.2, very close to "1111" NdFeAsO$_{1-x}$F$_x$ compound (4~6) near $T_c$ [33,34]. Several typical $\Gamma(H_{c2})$ values of iron based superconducting single crystals are listed in Table I for comparison. Such relative large anisotropy may result from the large distance $d$ (~1.035 nm) between the adjacent FeAs layers[24,39]. In addition, the large anisotropy values may also be affected by the transport properties of the barrier layer, such as Ca atom and unique As-As chain layers[24]. However, in order to determine the upper critical field and clarify the evolution of $\Gamma$ in the low temperature region, higher field experiments are needed in the future.

Now let us turn to the normal state properties of this 112-type IBS. Hall resistivity $\rho_{xy}$ at several temperatures for the $Ca_{0.8}La_{0.2}Fe_{0.98}Co_{0.02}As_2$ single crystal was shown in the inset of Fig.6. $\rho_{xy}$ shows a good linearity versus magnetic field up to 9 T. Temperature dependence of Hall coefficient $R_H(T)$ determined through $R_H = \rho_{xy}/H$ is presented in Fig.6. $R_H$ is negative over the whole temperature, suggesting the dominance of electron carriers. $R_H(T)$ shows strong and nonmonotonic temperature dependence. Specially, $R_H$ decreases initially with decreasing $T$, while it increases rapidly below $T_H \sim 175$ K. Similar behaviors were also observed in $(Li_{0.84}Fe_{0.16})OHFe_{0.98}Se$ system, which was attributed to the reduction of hole concentration[40].

In standard metals, the MR exhibits a $H^2$ dependence in the weak-field limit. Here we define $MR = \frac{\Delta\rho}{\rho(0)} = \frac{\rho(H)-\rho(0)}{\rho(0)}$, where $\rho(H)$ is the resistivity at a magnetic field $H$ and $\rho(0)$ is that at zero field. It can be seen clearly from Fig. 7(a) that the MR follows the quadratic field-dependence up to 9 T. The temperature dependence of MR under 9 T was plotted in Fig.7 (b). It is observed that the MR increases with decreasing $T$ before saturating at $T<100$ K, similar to the results of $(Ba,K)Fe_2As_2$[41]. According to Boltzmann transport theory, the $T$ dependence of MR can be scaled by the expression $\frac{\Delta\rho}{\rho(0)} = F(H\tau)$, known as the Kohler's law[42], where $\tau$ is the electron relaxation time. Since the resistivity $\rho(0)$ is related to the relaxation $\tau$ by $\rho(0) = \frac{m^*}{ne^2\tau}$, where $m^*$ is the effective mass and $n$ is the carrier density. Then Kohler's rule can be rewritten as

$$\frac{\Delta\rho}{\rho(0)} = F(H\tau) = F(\frac{H}{\rho(0)}\frac{m^*}{ne^2}) \qquad (1)$$

If the factor $\frac{m^*}{ne^2}$ is not sensitive to temperature, the Kohler's rule can be simplified as the

common form

$$\frac{\Delta\rho}{\rho(0)} = f(\frac{H}{\rho(0)}) \qquad (2)$$

Generally, the violation of Kohler's rule is believed to arise from the change of carrier number with $T$ or from the fact that the form of the anisotropic electron scattering rates changes with $T$[43]. Considering the parabolic dependence of the magnetoresistance in our compound, we plotted the MR against $(H/\rho(0))^2$ as shown in Fig.7 (c). It can be seen that the MR measured at different temperatures do not overlap and deviate from the simple Kohler's rule below a characteristic temperature $T_K$=175 K. As can be seen in the inset of Fig.7 (c), only the data at high temperature region (175, 200, 225 and 250 K) obey the simple Kohler's rule. In Fig.7 (d) we plot the MR as a function of $(\mu_0 H \cdot R_H / \rho(0))^2$ below 175 K with considering the temperature dependence of carrier density, and the MR data at different temperatures collapses into one curve. Based on our results, it seems that only one electron band dominates the normal state transport properties in our samples, which results in linear magnetic field dependence of $\rho_{xy}$. However, the carrier density of this dominating band changes with temperature below $T_K$, which is possibly caused by the multiband effect and some electron state phase transition at $T_K$. Interestingly, $T_K$ is well in accordance with $T_H$ as mentioned above, which suggests that the anomalous behaviors of the Hall effect and the magnetoresistance may have the same origin. Moreover, it has been proposed that the standard Kohler's law is not valid anymore when a pronounced anisotropic scattering rate is induced and the MR can be scaled by the Hall angle. The new scaling relation, $\frac{\Delta\rho}{\rho(0)} \propto \tan\Theta_H^2 = (\rho_{xy}/\rho_{xx})^2$ (modified Kohler's rule), which has shown to describe well the MR in some strongly correlated materials, such as high $T_c$ cuprates [44] and heavy fermion compounds[45]. Note that the linearity of $\rho_{xy}$ versus $H$, $(\rho_{xy}/\rho_{xx})^2 = (H \cdot R_H / \rho(0))^2$, the MR is also scaled well by the $(H \cdot R_H / \rho(0))^2$ as plotted in Fig.7(d). For optimally doped BaFe$_2$(As,P)$_2$ and Ba(Fe,Ru)$_2$As$_2$ systems, both violation of Kohler's rule and Hall angle scaling in normal state magnetoresistance were also observed and they were attributed to the anisotropic scattering rate due to the AFM scattering in the vicinity of the AFM phase[11,12]. Hence, the violation of the Kohler's rule in $Ca_{0.8}La_{0.2}Fe_{0.98}Co_{0.02}As_2$ may also come from the exotic anisotropic scattering inherent in the system.

## 4. Summary

In summary, X-ray diffraction, magnetization, resistivity, and transport measurements were performed on high quality $Ca_{0.8}La_{0.2}Fe_{0.98}Co_{0.02}As_2$ single crystals. The magnetization measurement revealed a second-peak effect, which was observed for the first time in this system. The value of $J_c$ is over $2\times10^6$ A/cm$^2$ at 5 K (self-field), making it promising for practical applications. The upper critical field anisotropy was obtained by measuring electrical resistivity at different magnetic fields and angles. The normal state transport shows a strong but nonmonotonic $T$ dependence of the Hall coefficient $R_H$ with a minimum value at ~175 K, below which the Kohler's rule was also violated. The violation of Kohler's rule implies that the $T$-variation of $R_H$ may be responsible for this behavior and suggests the strong temperature dependence of carriers or an exotic anisotropic scattering. The anomalous features of transport properties observed in this study however invoke further investigations in the future.

## Acknowledgements

This work was supported by the National Natural Science Foundation of China (Grant No. NSFC-U1432135, 11474080), by Natural Science Foundation of Jiangsu Province of China (Grant No.BK20141337), by Distinguished Young Scientist Funds of Zhejiang Province (LR14A040001), by the Scientific Innovation Research Foundation of College Graduate in Jiangsu Province (KYZZ0063), and by the Scientific Research Foundation of Graduate School of Southeast University.

**Table I** Superconducting parameters ($T_c$, critical current density $J_c$ and upper critical field anisotropy $\Gamma(H_{c2})$) near $T_c$ of several typical iron based superconducting single crystals.

| Sample | $Ca_{0.8}La_{0.2}Fe_{0.98}Co_{0.02}As_2$ (112-type, this work) | $Ba_{0.6}K_{0.4}Fe_2As_2$ (122-type)[28,37] | $Fe_{0.6}Te_{0.4}Se$ (11-type)[29] | $NdFeAsO_{0.7}F_{0.3}$ (1111-type)[33] | LiFeAs (111-type)[30,38] | $Ca_{10}(Pt_{4-\delta}As_8)(Fe_{2-x}Pt_xAs_2)_5$ (1048-type)[31,39] |
|---|---|---|---|---|---|---|
| $T_c$ | 37.8 K | 36.5 K | 14 K | 51.5 K | 16.5 K | 33 K |
| $J_c$ (5 K, 0 T) | $2 \times 10^6$ A/cm$^2$ | $3 \times 10^6$ A/cm$^2$ | $10^5$ A/cm$^2$ | -- | $1.2 \times 10^5$ A/cm$^2$ | $0.8 \times 10^5$ A/cm$^2$ |
| $\Gamma(H_{c2})$ | 2.4~5.2 | 1~2 | 1~2 | 4~6 | 1.5~2.4 | 7~10 |

# Figure Captions

Figure 1. Temperature dependence of the electrical resistivity of $Ca_{0.8}La_{0.2}Fe_{0.98}Co_{0.02}As_2$ single crystal. Top inset: the x-ray diffraction of $Ca_{0.8}La_{0.2}Fe_{0.98}Co_{0.02}As_2$ single crystal. Bottom inset: temperature dependence of ZFC and FC magnetization at 20 Oe for $Ca_{0.8}La_{0.2}Fe_{0.98}Co_{0.02}As_2$.

Figure 2. The magnetic hysteresis loops (a) and the magnetic field dependence of the critical current density $J_c$ (b) calculated from the Bean model at several temperatures ranging from 5 K to 36 K.

Figure 3. Magnetic field dependence of in-plane resistivity for (a) $H\|ab$ and (b) $H\|c$. (c) Temperature dependence of upper critical field for $H\|c$ and $H\|ab$ obtained from the 50% normal state resistivity.

Figure 4. Main panels: scaling of the resistivity versus $H\sqrt{\sin^2(\theta)+\Gamma^2\cos^2(\theta)}$ at (a) T=34 K, (b) T=35 K, (c) T=36 K, and (d) T=37 K in different magnetic fields. Insets: the angular dependence of magnetoresistance in magnetic fields up to 9 T corresponding to respective

temperature.

Figure 5. The temperature-dependent $\Gamma$ for $Ca_{0.8}La_{0.2}Fe_{0.98}Co_{0.02}As_2$ single crystal. The red filled circles represent the anisotropic $\Gamma$ determined using the AGL theory.

Figure 6. Temperature dependence of the Hall coefficient $R_H$ for $Ca_{0.8}La_{0.2}Fe_{0.98}Co_{0.02}As_2$ single crystal. Inset: The Hall resistivity $\rho_{xy}$ versus the magnetic field at several temperatures.

Figure 7. (a) Field dependence of MR at several temperatures. (b) Temperature dependence of MR at 9 T. (c) Kohler's scaling of MR at various temperatures. Inset: Kohler's scaling in the high temperature region from 175 K to 250 K. (d) MR plotted as a function of $(\mu_0 H \cdot R_H / \rho(0))^2$.

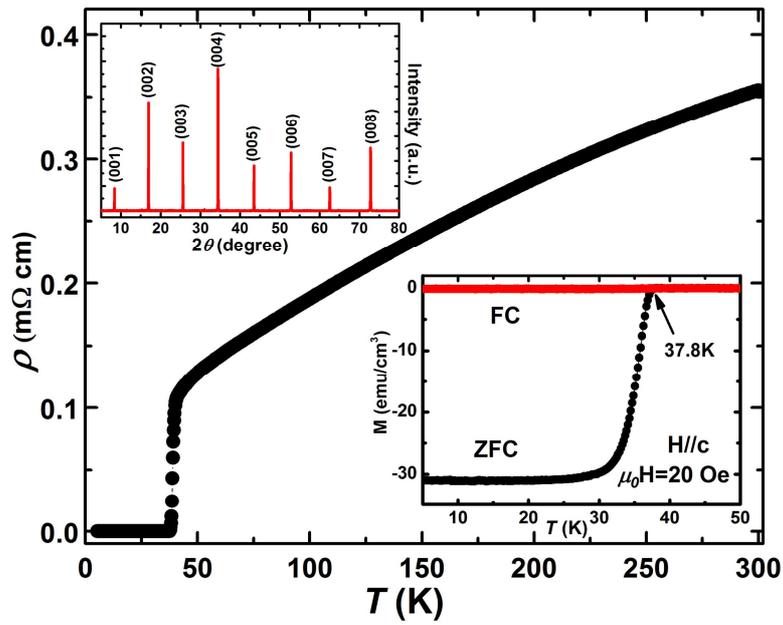

Figure 1.

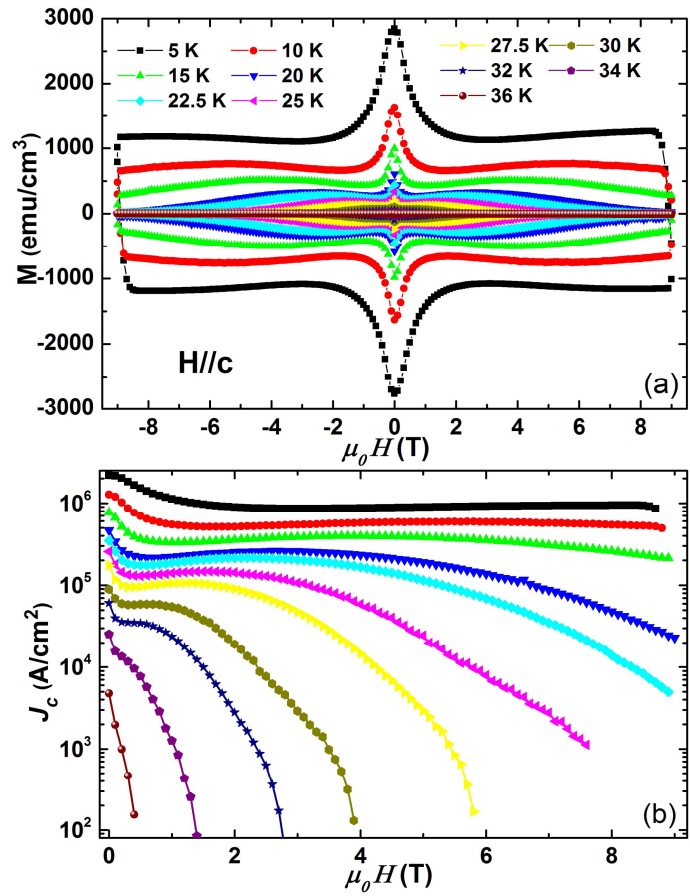

**Figure 2.**

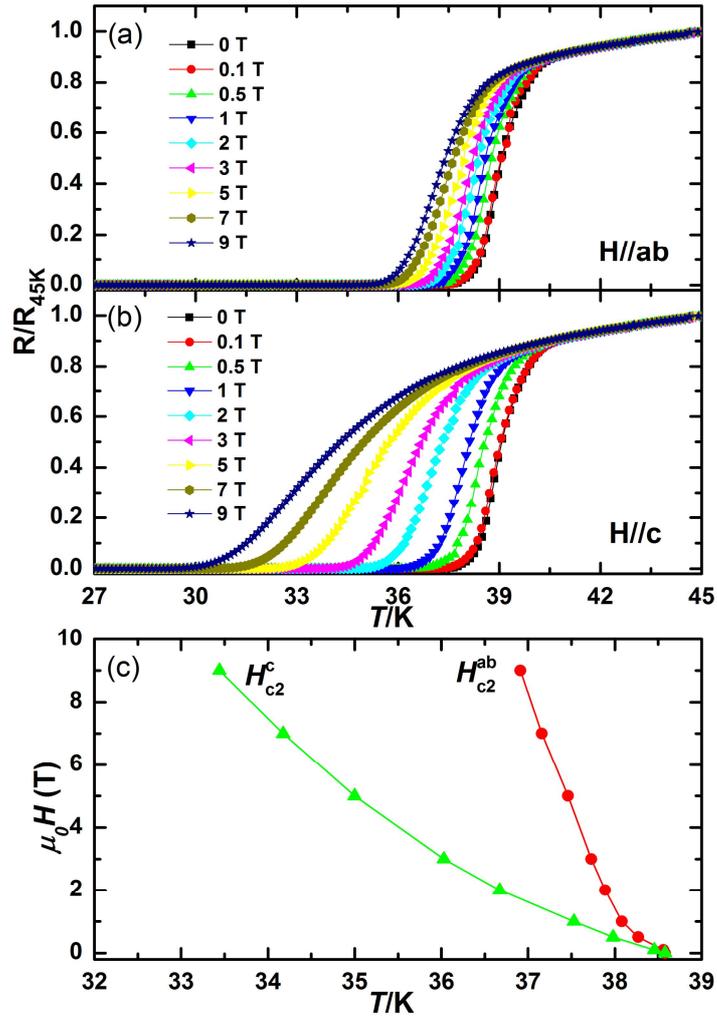

Figure 3.

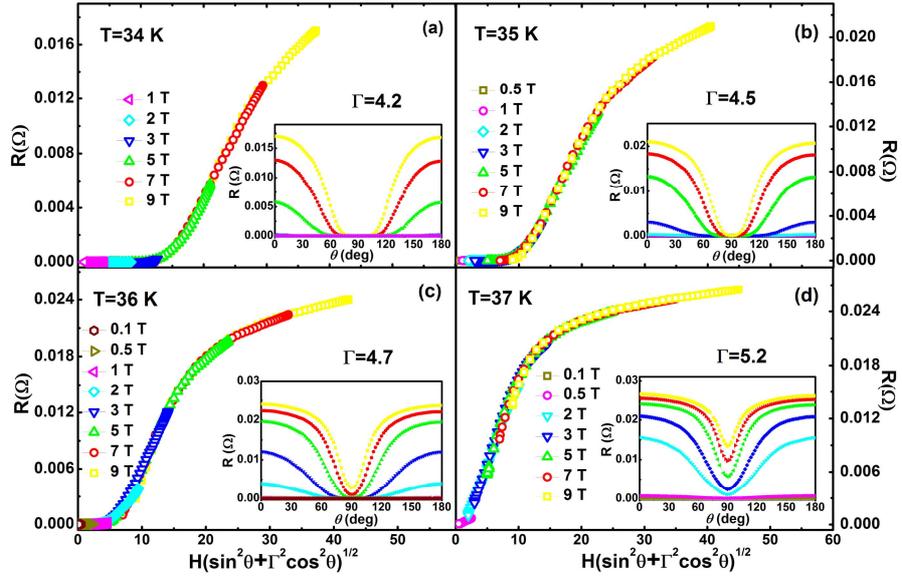

**Figure 4.**

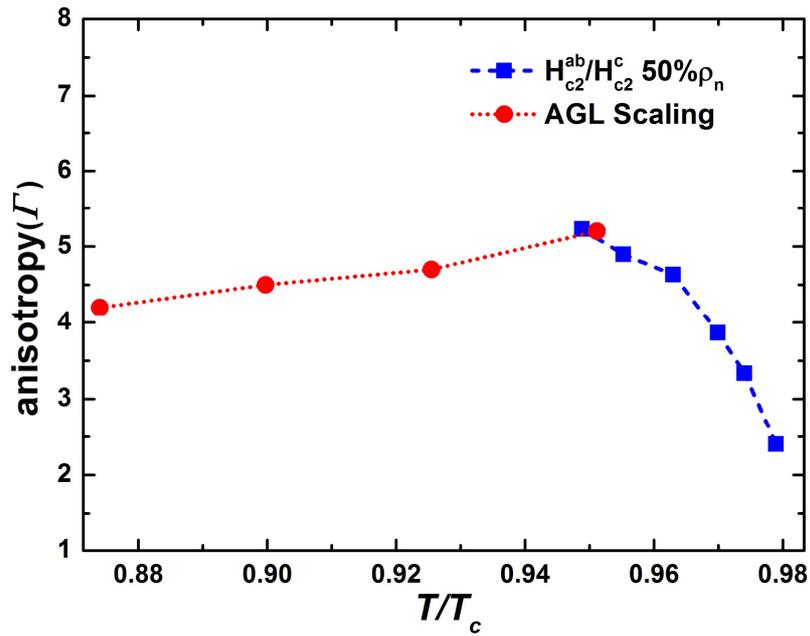

**Figure 5.**

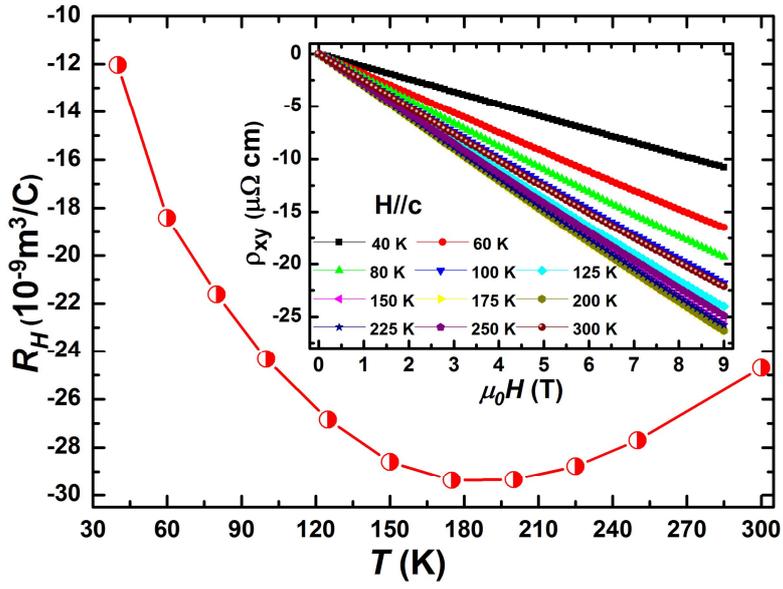

**Figure 6.**

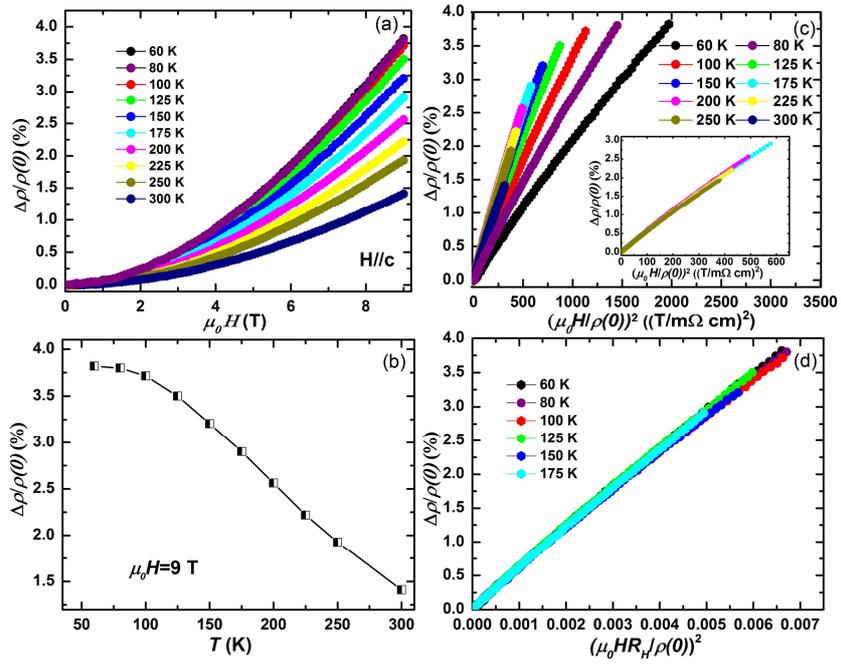

**Figure 7.**